
\magnification=1200
\input definit.tex
\baselineskip=22truept
\centerline{\bf{Efimov Effect Revisited with Inclusion of Distortions}}

\ \

\bigskip \centerline{B.G.Giraud}

\centerline{Service de Physique Th\'eorique, DSM, CE Saclay,
F-91191 Gif/Yvette, France}

\bigskip \centerline{Yukap Hahn}

\centerline{Physics Dept, University of Connecticut, Storrs, CT, 06269, USA}

\ \

\bigskip \noindent
{\bf Abstract:} An elementary proof of the 3-body Efimov effect is provided
in the case of a separable 2-body potential which binds at zero energy
a light particle to a heavy one. The proof proceeds by two steps, namely
{\it i)} a projection of the Hamiltonian in a subspace and the observation that
the projected Hamiltonian generates an arbitrarily large number of bound
states, and {\it ii)} a use of the Hylleraas-Undheim theorem to recover the
unprojected Hamiltonian. The definition of the projectors we use can include
mean field distortions.

\ \

\bigskip \centerline{1. Introduction}

\bigskip
The Efimov effect$^{1)}$ consists in the claim that those 2-body forces
which marginally bind pairs of particles in the 2-body problem
induce an arbitrarily large number of bound states for the 3-body problem.
This interesting phenomenon has been studied by many authors$^{2-5)},$
often in the framework of Faddeev equations, and, in particular,
was validated by a formal proof$^{2)}.$ An investigation of
the same effect for the 4-body problem concluded$^{6)}$
that the effect does not exist in the case of identical bosons, but is present
if three heavy particles meet with a fourth lighter particle$^{7,8)}.$

\medskip
In this work, we follow the argument of Fonseca, Redish and Shanley$^{3,4)}$
and give again a proof of the effect, by means of a very elementary
derivation. Symmetrization constraints between the three particles are strongly
relaxed, two particles being chosen as heavy and the third particle
as light. We do not use Faddeev equations, but
rather a Born-Oppenheimer (BO) approach. This is the subject of Section 2. Then
Section 3 reinterprets the BO method as a generator coordinate method, thus
as a projection into a trial subspace. We complete the proof by
{\it i)} estimating the non adiabaticity corrections and {\it ii)}
considering the residual coupling effects of the full Hamiltonian.
Self consistent distortions of
pair wave functions due to the presence of the third particle are
discussed in Section 4.
Finally we provide a discussion and conclusion in Section 5. 

\ \

\bigskip
\centerline{2. Elementary Proof}

\bigskip
We consider the familiar Jacobi coordinates $\vec x$ for pair (12)
and $\vec y$ for particle 3 with respect to the center of mass of (12). Then we
consider the 3-body Hamiltonian
$$
H \equiv
- \hbar^2 \Delta_{\vec x} / M + w(\vec x) - \hbar^2\Delta_{\vec y} / (2\mu)
+ \lambda [v(\vec y-\vec x/2)+v(\vec y+\vec x/2)],\ \ \mu=m/(1+\varepsilon/2),
\eqno(2.1)
$$
where, for the sake
of simplicity, the three particles are spinless, the masses $M_1$ and $M_2$
of particles 1 and 2 are equal, $M_1=M_2=M,$ and the third mass $m$ is much
lighter, $\varepsilon \equiv m/M<<1.$

\medskip
Both potential functions $v$ and $w$ are assumed to have a short or a finite
range, but their exact forms, local or even non local,
are otherwise irrelevant. For the sake of simplicity, we assume
that $w$ is too weak to create bound states between the two heavy
particles. Hence the continuum in the three-body problem could
display two sorts of cuts at most, namely {\it \i)}the three-body continuum
at energy zero, naturally, and {\it ii)}two body-continuum(s) if the
interaction $v$ were able to bind a $(m+M)$ pair. Actually, the strength
parameter $\lambda$ has that critical value which
makes {\it semi}-positive definite the 2-body Hamiltonian
$$
h_0 \equiv - \hbar^2 \Delta_{\vec z}/(2\mu) + \lambda v(\vec z). \eqno(2.2)
$$
Here the degree of freedom $\vec z$ may be
either {\it i)} $\vec z = \vec y$ or {\it ii)}
$\vec z = \vec y \pm \vec x/2,$
but it will be understood that $h_0$ acts only upon the degree of freedom
$\vec y.$ The fact that $h_0$
generates a ``zero energy ground state'' does not depend on $\vec x,$
which plays for $h_0$ the r\^ole of a simple parameter,
just defining the center of the critical potential $\lambda v.$

\medskip
It is noted that the critical value
of $\lambda$ is linked here, from Eq.(2.2), to the value of
the reduced mass $\mu$ rather than the slightly smaller reduced
mass of a pair, $mM/(M+m)=m/(1+\varepsilon).$
With such a value of $\lambda,$ critical for $\mu,$ no bound
state is induced
for the 2-body problem with $m/(1+\varepsilon).$
The 2-body continuum in the 3-body
problem starts at the same energy ($E=0,$ indeed) as the 3-body continuum.
At most a resonance
may occur in the 2-body channel at a slightly positive energy $E_r.$

\medskip
For the sake of simplicity, the argument which follows assumes that $v$ is
a scalar, non local, rank one, separable potential$^{4)},$
$$
<\vec p\ '|v|\vec p>= - f(p') f(p), \eqno (2.3)
$$
where $\vec p$ is the momentum conjugate to $\vec z,$ and
$p,p'$ are the lengths of $\vec p, \vec p\ ',$ respectively.
The well-known binding condition for separable potentials then reads,
at the critical value of $\lambda,$
$$
1/\lambda = \int d \vec p \ [f(\vec p)]^2/p^2, \eqno(2.4)
$$
Moreover
an exactly soluble model is obtained if we choose the following
form factor $f,$
$$
f(p)=1/[\pi(p^2+\gamma^2)], \eqno(2.5)
$$
where $1/\gamma$ defines any suitable short range in coordinate space.
With Eq.(2.5), the condition, Eq.(2.4), for a critical value of $\lambda$
becomes
$$
\lambda=\gamma^3. \eqno(2.6)
$$
In the following we set $\hbar=1$ and $\hbar^2/(2\mu)=1,$ for a simpler system
of units. We
also choose $\vec p$ to be the momentum conjugate to $\vec y,$ and introduce
the translation operator $T=\exp(i \vec p \cdot \vec x /2)$
and its inverse
$T^{-1}=\exp(-i \vec p \cdot \vec x /2).$
Then that part $h_f$ of $H$ which acts upon $\vec y$ reads,
$$
h_f = p^2 + \lambda ( T v T^{-1} + T^{-1} v T )
=p^2 - \lambda ( T |f><f| T^{-1} + T^{-1} |f><f| T ). \eqno(2.7)
$$
Parametrized by $\vec x,$ this Hamiltonian $h_f$
drives the BO dynamics of the fast, light particle.
When compared with the semi-positive definite $h_0,$ this Hamiltonian
$h_f$ contains one additional attractive potential, and thus
induces a truly bound state (square integrable)
$\chi_{\vec x}(\vec y)$ at a definitely negative energy $\eta_{\vec x}.$
Since threshold singularities of a square root
nature with respect to $\eta$ are expected, we define
such a square root $\omega$ by $\eta_{\vec x}=-[\omega_{\vec x}]^2.$
Naturally $\omega_{\vec x} \rightarrow 0$ when $ x \rightarrow \infty.$ The
subscript $\vec x$ will be omitted in the following, unless it is essential.

\medskip
The wave function $\chi$ is easily obtained in momentum representation,
according to
$$
\chi(\vec p) = \lambda <f|T|\chi> 2 \cos(\vec p \cdot \vec x/2)\
f(\vec p)/(p^2+\omega^2),
\eqno(2.8)
$$
where we take advantage of the symmetry of the ground state,
$<f|T^{-1}|\chi>=<f|T|\chi>.$ Projecting Eq.(2.8) against $<f|T,$ we obtain
the equation which solves for the binding energy,
$$
1/\lambda=I(\omega,\vec x)+I(\omega,0),\eqno(2.9)
$$
with
$$
I(\omega,\vec x) \equiv <f|(p^2+\omega^2)^{-1}T^2|f>. \eqno(2.10)
$$
We note incidentally that the condition
for the value of $\lambda$ to be critical reads
$$
1/\lambda=I(0,0). \eqno(2.11)
$$

With the choice of $f(\vec p)$ as a scalar $f(p),$ the integral $I,$ Eq.(2.10),
does not depend on the orientation $\hat x$ of $\vec x,$
but only on its length $x.$ More precisely, it becomes
$$
I(\omega,x) = 4\pi/x \int_0^{\infty}dp\ p\ \sin(px)\
[f(p)]^2/(p^2+\omega^2).
\eqno(2.12)
$$
For the soluble model provided by the choice, Eq.(2.5), a straighforward
contour integration gives
$$
\eqalign
{
I(\omega,x)=&2/(\pi x)\int_{-\infty}^{\infty}dp\ p\ \sin(px)
/[(p^2+\omega^2)(p^2+\gamma^2)^2] \cr
=&[2 \gamma \exp(-\omega x)+(\omega^2 x - \gamma^2 x - 2 \gamma)
\exp(-\gamma x)] /[\gamma x (\gamma^2-\omega^2)^2]. \cr
} \eqno(2.13)
$$
It is easy to obtain the value of $I$ for $x=0,$
$$
I_0(\omega) \equiv I(\omega,0)=1/[\gamma(\gamma+\omega)^2]. \eqno(2.14)
$$
and then, for Eqs.(2.11,6), the number $I_{00} \equiv I(0,0)=1/\gamma^3.$

\medskip
If $\omega x$ has a finite limit $c$ when $x \rightarrow \infty ,$
it is also easy to find that $I(\omega,x)$ boils down to
$$
I_{\infty} = {2 \exp(-\omega x) \over x (\gamma^2 - \omega^2)^2 }.
\eqno(2.15)
$$
This simplification is also useful for the calculation of derivatives of $I,$
because the contributions of the discarded term, proportional to
$\exp(-\gamma x),$ are negligible for derivatives as well.
Let us test this conjectured
finite limit of $\omega x,$ by means of an ansatz
$\omega=c/x+d/x^2+{\cal O}(1/x^3)$
for Eq.(2.9). We obtain the condition
$$
{1 \over \gamma^3}=
{2\exp(-c-d/x) \over x \left[\gamma^2 - (c+d/x)^2/x^2\right]^2 }
+ { 1 \over \gamma \left[ \gamma + (c+d/x) / x \right]^2 } + {\cal O}(1/x^3).
\eqno(2.16)
$$
The ansatz is then consistent if $c$ and $d$ obey the conditions
$$
\exp(-c) = c,\ \ d = 3 c^2 / [ 2 \gamma ( 1 + c ) ], \eqno(2.17)
$$
hence $c \simeq 0.5671,\ d \simeq 0.3079/\gamma.$
We conclude that the ``fast'' BO Hamiltonian$^{4,5)}$
$h_f$ binds the light particle at energy $\eta(x)=-c^2/x^2+{\cal O}(1/x^3)$
when $x \rightarrow \infty.$

\medskip
For a better understanding of this $1/x^2$ binding,
a thorough examination of Eq.(2.9) is in order.
We rephrase it as
$$
1/\lambda - I(\omega,0) = I(\omega,x), \eqno(2.18)
$$
and notice on one hand
that, because of the Fourier transform introduced explicitly
by the translation $T^2$ in the definition of $I,$ Eq.(2.10), we can
take advantage of Lebesgue's theorem to predict a $1/x$ factor in the
right-hand side of Eq.(2.18) when $x \rightarrow \infty.$
On the other hand, the left-hand side of Eq.(2.18), once it is written
as a function of $\omega$ rather than $\eta,$ is a {\it regular} function
near $\omega=0.$ Since $1/\lambda=I(0,0),$ the difference $I(0,0)-I(\omega,0)$
is proportional to first order with respect to $\omega.$ Hence the
possibility of a connection $\omega \propto 1/x.$ Had $\lambda$ been
overcritical, inducing an actual negative eigenvalue
$-\omega_0^2,\ \omega_0 \ne 0,$
then $1/\lambda=I(\omega_0,0),$ and one might have
tried a connection $\omega-\omega_0 \propto 1/x,$ but failed because
of the resulting decay of $\exp(-\omega_0 x)$ in the right-hand side.
This shows the strict importance, for the
Efimov argument, of the condition that the value $\lambda$ be exactly
critical.

\medskip
It is now trivial to consider the ``slow'' BO Hamiltonian,
which drives the heavy degree of freedom $\vec x,$
$$
h_s = - \Delta_{\vec x} / M + w(\vec x) + \eta(x). \eqno(2.19)
$$
The effective potential $\eta(x)$ induced by the binding of the light particle
has the expected long range behavior, $\eta \propto -1/x^2,$ that justifies
the Efimov prediction of an infinite number of bound states
for the 3-body system. It is interesting to point out, incidentally, that the
mass ratio $\varepsilon$ appears strictly
nowhere in this derivation of $\eta.$ Hence, if the decoupling
of the full Schroedinger equation into the two separate steps described by
$h_f,h_s$ can be justified by
any other argument than the Born-Oppenheimer one,
it can be claimed that the $-1/x^2$ nature of the induced potential is valid
also for particles with similar masses.

\medskip
We can also offer a qualitative argument for a generalisation of this
$-1/x^2$ nature in the case of potentials $v$ other than separable. Let us
assume that $v$ is semi-negative definite and let $\rho(\omega)$
be the highest eigenvalue of the operator
$R(\omega) \equiv (-v)^{1/2}(p^2+\omega^2)^{-1}(-v)^{1/2}.$
In the same way, let $\sigma(\omega,x)$ be the highest eigenvalue of
$S(\omega,x) \equiv
(-v-T^2vT^{-2})^{1/2}(p^2+\omega^2)^{-1}(-v-T^2vT^{-2})^{1/2}.$
The fact that $\lambda$
is critical is expressed by the condition $1/\lambda=\rho(0).$ Then the
binding energy in presence of the
two potentials $TvT^{-1}$ and $T^{-1}vT$ is obtained
when solving for $\omega$ the ``spectral'' condition
$\rho(0)=\sigma(\omega,x).$
If, when $x \rightarrow \infty,$ a property of the form
$\sigma(\omega,x)=\rho(\omega)+\tau(\omega x)/x$ occurs, where $\tau$
is a suitable function depending on the product $\omega x$ only,
then the ``spectral'' condition which provides $\omega$
amounts to $\rho(0)-\rho(\omega)=\tau/x.$ It is also reasonable
to assume that $\rho(0)-\rho(\omega)={\cal O}(\omega)$ when
$\omega \rightarrow 0.$ Then the whole scheme discussed for Eq.(2.18)
is recovered, and thus $\omega \propto 1/x.$

\medskip
In summary for this section, elementary arguments
indicate that if the distance
$x$ between two particles is frozen, and if a third particle is ``bound
at energy zero'' to, e.g., the first particle, the presence of the second
particle will induce true binding of the third particle,
at an energy $\eta \propto -1/x^2$ when $x \rightarrow \infty.$ This $\eta$
will in turn behave as a long range potential, efficient for creating
many bound states of the pair made by the first and second particles.

\ \

\bigskip
\centerline{3. Born-Oppenheimer Approximation and Generator Coordinate Method}

\bigskip
Let $\varphi$ be a bound eigenstate of $h_s,$ and $E$
the corresponding eigenvalue. For the sake of simplicity,
we assume in the following that $w$ is a scalar, like $\eta$ and $v,$ and
we then omit vector notations whenever possible. The function
$\varphi$ is a scalar ($s$ state).
Although $\chi_{\vec x}$ is not a scalar, it transforms very simply under
any rotation $\cal R,$ namely
${\cal R}(\chi_{\vec x})=\chi_{{\cal R}(\vec x)}$.
As stated already, the vector nature of $\vec x$ as a label can often
be understood, and/or shortened into a simple scalar label $x.$

\medskip
The well-known BO ansatz for the 3-body state $\psi(x,y)$
corresponding to $E$ reads,
$$
\psi(x,y)=\varphi(x) \chi_x(y). \eqno(3.1)
$$
This may be written as well under the form,
$$
\psi(x,y)=\int d\xi\ \varphi(\xi)\ \delta(x-\xi)\chi_{\xi}(y), \eqno(3.2)
$$
which is nothing but a generator coordinate$^{10-12)}$ expansion on a basis
$$
\phi_{\xi}(x,y)=\delta(x-\xi)\chi_{\xi}(y), \eqno(3.3)
$$
of states $\phi_{\xi}$ parametrized by the continuous label $\xi.$ The ``slow''
wave function $\varphi(x)$ is thus reinterpreted as a
mixture amplitude $\varphi(\xi)$ with respect to the
generator coordinate $\xi.$

\medskip
If this reinterpretation can be shown to be consistent, we will
thus obtain that any
BO eigenvalue $E$ is an eigenvalue of the projection $PHP$
of $H$ in that subspace spanned by the wave functions $\phi_{\xi}.$
This will reinforce the BO result, namely that
there are an infinity of such negative eigenvalues $E.$
Indeed, we can then take advantage of
the Hylleraas-Undheim (HU) theorem$^{9)}$, which
we summarize by the short statement ``the unprojected Hamiltonian
has more (or at least as many) bound states than (as) its projection
into any subspace ''.

\medskip
That is, any exact wave function $\Psi$ may be written as
$$
\Psi \ = \ P \Psi \ + \ Q \Psi, \eqno(3.4)
$$
where $Q$ is a projector orthogonal to $P.$
Then, the effect of the
added $Q$-term modifies the spectrum of $PHP$ in such a way that
the correct number $N$ of bound states satisfies the inequality
$$
N_P \ \  \le \ \  N_t \ \ \ \le \ \ N \ \ .  \eqno (3.5)
$$
Here $N$ is the exact number of bound states of the original $H,$ but
we have also
inserted an intermediate number $N_t$ of
bound states in Eq.(3.5) to include
the situation where $Q$ is not the full complement $1-P$ of $P,$ but only part
of it, for a variational theory with trial states $\Psi_t$
in an enlarged subspace.
In many of the earlier treatments of the Efimov effect in terms of models
and simple ansatz on the wave functions, the effect of the $Q$-component
of the wave function was not estimated, which made the proof
incomplete.  The HU theorem
for the bound states
may be re-stated as
a deepening influence of the effective $Q$-space potential $PV\ G^Q\ VP < 0$
where $G^Q$ is the Green's function in the $Q$-space.
Therefore, all the bound states produced by $PHP$ will be uniformly pushed
down in energy, and possibly additional bound states are created by
the $Q$-component.

\medskip
We now proceed by proving that the slow Hamiltonian, Eq.(2.19), is formally
equivalent to a Griffin-Hill-Wheeler (GHW) kernel
$H(\xi',\xi) \equiv <\phi_{\xi'}|H|\phi_{\xi}>,$ hence a representation of the
projection $PHP$ of $H$ in the generator subspace spanned by the states
$\phi_{\xi}$. For this, we first notice
that such states $\phi_{\xi}$ are not orthogonal
to one another if one measures their scalar product by
$<\chi_{\xi'}|\chi_{\xi}>,$ namely an integration upon $y$
only, but become trivially orthogonal if an integration upon $x$ is also
included. Indeed such states are strictly localized in $x$-space and
can be normalized according to $<\phi_{\xi'}|\phi_{\xi'}>=\delta(\xi'-\xi). $
Nothing prevents us from normalizing these states to unity in the
$y$-space, since they are strictly bound eigenstates of the fast Hamiltonian
$h_f,$ Eq.(2.7),
$$
\int dy\ \left[\chi_{\xi}(y)\right]^2=1. \eqno(3.6)
$$
We find incidentally, as a consequence of Eq.(3.6), that
$$
\int dy\ \chi_{\xi}(y)\ \left[\nabla_{\xi}\ \chi_{\xi}(y)\right]=
{1 \over 2} \nabla_{\xi} \left(\int dy\ [\chi_{\xi}(y)]^2\right)=0, \eqno(3.7)
$$
and furthermore, as a consequence of Eq.(3.7), that
$$
u(\xi) \equiv \int dy\ \chi_{\xi}(y)\ [-\Delta_{\xi}\ \chi_{\xi}(y)]=
\int dy\ [\nabla_{\xi}\ \chi_{\xi}(y)]^2\ >0.
\eqno(3.8)
$$
The GHW kernel thus reads
$$
\eqalign
{
<\phi_{\xi'}|H|\phi_{\xi}>=&
<\phi_{\xi'}|[-\Delta_x/M+w(x)+h_f(x,y)]|\phi_{\xi}> \cr
=& \delta(\xi'-\xi)
\left({2\varepsilon \over 1+\varepsilon/2}
[-\Delta_{\xi}+u(\xi)]+w(\xi)+\eta(\xi)\right). \cr
}
\eqno(3.9)
$$
Except for a correction proportional to $u(\xi),$ this is nothing
but $h_s$ in coordinate representation. The GHW equation,
$$
\int d\xi\ <\phi_{\xi'}|(H-E)|\phi_{\xi}>\ \varphi(\xi) = 0, \eqno(3.10)
$$
and the slow BO equation,
$$
\left(-{2\varepsilon \over 1+\varepsilon/2}\Delta_x+w(x)+\eta(x)-E\right)
\varphi(x)=0, \eqno(3.11)
$$
are thus equivalent if this correction
brought by $u,$ a repulsive effective potential, can be shown to be negligible.
For that purpose we restate Eq.(2.8) in the form
$$
|\chi>=(< \zeta | \zeta >)^{-1/2}|\zeta>, \eqno(3.12)
$$
with
$$
|\zeta>=(p^2+\omega^2)^{-1}(T+T^{-1})|f>,
\ \ <\zeta|\zeta>=2<f|(p^2+\omega^2)^{-2}(1+T^2)|f>.
\eqno(3.13)
$$
In particular, when $x \rightarrow \infty,$
$$
\eqalign {
 \omega <\zeta|\zeta> & =
-{\partial \over \partial\omega}[I_0(\omega)+I_{\infty}(\omega,x)] \cr
& = {2 \over \gamma (\gamma+\omega)^3 }
  + { 2 \exp(-\omega x) \over  (\gamma^2-\omega^2)^2}
  - {8 \omega \exp(-\omega x) \over x (\gamma^2-\omega^2)^3 }. \cr
} \eqno(3.14)
$$
The right-hand side of Eq.(3.14) has a finite limit
$ \ell = 2 [1+\exp(-c)] / \gamma^4 $ when $x \rightarrow \infty,$ hence
$<\zeta|\zeta>=2(1+c)x/(c\gamma^4) + {\cal O}(x^0)$
when $x \rightarrow \infty.$

\medskip
Let us denote by $|\vec \chi'>,|\vec \zeta'>$ the full gradients of
$|\chi>,|\zeta>,$ respectively, with respect to $\vec x,$
including the dependence of $\omega$ on $x.$ It is easy to find that
$$
|\vec \zeta'>=
{i \vec p \over 2} (p^2+\omega^2)^{-1} (T-T^{-1}) |f>
- 2 \omega (p^2+\omega^2)^{-2}(T+T^{-1}) |f> \nabla_{\vec x} \ \omega,
\eqno(3.15)
$$
where we know that
$\nabla_{\vec x}\ \omega = -(c+2d/x) \hat x / x^2 + {\cal O}(1/x^4)
= -(c+2d/x) \vec x / x^3 + {\cal O}(1/x^4).$ Here
$\hat x$ is the unit vector defining the direction of $\vec x,$
namely $\hat x=\vec x /x.$ Then we have to investigate the behavior of
$$
u = <\vec \chi'|\vec \chi'>=(<\zeta|\zeta>)^{-1}
<\vec \zeta'| \left(1-{|\zeta><\zeta| \over <\zeta|\zeta>} \right)|\vec \zeta'>
, \eqno(3.16)
$$
when $x \rightarrow \infty.$

\medskip
{}From the linear asymptotic behaviour, $<\zeta|\zeta>\ \propto x,$
of the square norm of $\zeta,$  we deduce that the length of its
gradient, namely of $\ 2<\zeta|\vec \zeta'>,$ has a finite limit
when $x \rightarrow \infty.$
Indeed, from the derivative of the right-hand side of Eq.(3.14), we find
$$
\eqalign {
& <\zeta|\vec \zeta'> = \cr
& \left[ - { \exp(-\omega x) \over (\gamma^2-\omega^2)^2}
  + {4 \omega \exp(-\omega x) \over x (\gamma^2-\omega^2)^3 }
  + {4 \exp(-\omega x) \over x^2 (\gamma^2-\omega^2)^3 } \right] \hat x
  - \left[ {1 \over \omega^2 \gamma (\gamma+\omega)^3 }
  + {3 \over \omega \gamma (\gamma+\omega)^4 } \right]
 \nabla_{\vec x}\ \omega \cr
& - \left[ { x \exp(-\omega x) \over \omega (\gamma^2-\omega^2)^2 }
  + { \exp(-\omega x) \over \omega^2 (\gamma^2-\omega^2)^2}
  - { 8 \exp(-\omega x) \over  (\gamma^2-\omega^2)^3}
  + {24 \omega \exp(-\omega x) \over x (\gamma^2-\omega^2)^4 }
  \right]  \nabla_{\vec x}\ \omega, \cr
} \eqno(3.17)
$$
the limit of which is
$(1+c) \hat x/(c\gamma^4)=[\exp(c)+1] \hat x / \gamma^4.$
It can be concluded that the term $-(<\vec \zeta '|\zeta>/<\zeta|\zeta>)^2$
in the right hand side of Eq.(3.16) reinforces the potential
$\eta=-c^2/x^2$ by a contribution $-\varepsilon/[(2+\varepsilon)x^2].$

\medskip
We must now evaluate
$$
\eqalign {
& <\vec \zeta'|\vec \zeta'> = \cr
& {1 \over 2}<f|p^2(p^2+\omega^2)^{-2}(1-T^2)|f> -
4\omega (c+2d/x)
<f|\vec p \cdot \vec x(p^2+\omega^2)^{-3} \sin(\vec p \cdot \vec x)|f> /x^3
 \cr
& +8\omega^2(c+2d/x)^2<f|(p^2+\omega^2)^{-4}(1+T^2)|f>/x^4+{\cal O}(1/x^2).\cr
} \eqno(3.18)
$$
The matrix element of $\ \vec p \cdot \vec x \sin(\vec p \cdot \vec x) \ $
can be obtained as
$$
\eqalign {
& <f| \vec p \cdot \vec x \sin(\vec p \cdot \vec x) (p^2+\omega^2)^{-3} |f> =
{\cal I}m
<f| \vec p \cdot \vec x \exp (i \vec p \cdot \vec x) (p^2+\omega^2)^{-3} |f> =
\cr
& {\cal I}m
<f|\ \left[ -i {d \over d s} \exp(is \vec p \cdot \vec x) \right]_{s = 1}
(p^2+\omega^2)^{-3} \ |f>
= - \left( {d \over 2\omega d \omega} \right)^2 {d \over d s}\
I(\omega, s x)/2 \ |_{s=1}. \cr
} \eqno(3.19)
$$
Then a somewhat tedious calculation, which cannot be reported
here in detail, gives the contribution of $<\zeta'|\zeta'>$ to $u,$
$$
\eqalign {
& {2 \varepsilon <\zeta'|\zeta'> \over (1+\varepsilon/2) <\zeta|\zeta>} =
\left( { \varepsilon \over 2+\varepsilon } \right) \times \cr
& \left[ { c \gamma \over ( 1 + c ) x } +
{ 12 + 36 c + 45 c^2 + c^3 - 13 c^4 - 2 c^5
 \over 6 (1 +c)^3 x^2 } + {\cal O} (x^{-3}) \right] \simeq
 \left( { \varepsilon \over 2+\varepsilon } \right)
\left( { 0.36 \gamma \over x} + {1.98 \over x^2} \right). \cr
} \eqno(3.20)
$$
It turns out, as a matter of fact, that $<\zeta'|\zeta'>$ has a finite
limit $1/(2\gamma^3)$ when $x \rightarrow \infty.$ Combined with the result
already found, $<\zeta|\zeta>=2(1+c)x/(c\gamma^4) + {\cal O}(x^0),$ this
asymptotic behavior of $<\zeta'|\zeta'>$ induces a repulsive potential of
order $1/x,$ which seems to contradict the leading term $-c^2/x^2$
derived from the straight BO approximation.

\medskip
We notice however that this repulsive potential
$\varepsilon c \gamma/[(2+\varepsilon)(1+c)x]$, and also the minute
corrections brought by $u$ at order $1/x^2,$ are weighted by
$\varepsilon,$ naturally.
Hence the Efimov phenomenon is still present
for $\varepsilon$ sufficiently small, because a generic potential
of the form $\ \varepsilon/x-1/x^2\ $ generates
an arbitrarily large number of bound states.

\medskip
More precisely, if we collect all relevant terms contributing to Eqs.(3.9,10),
the radial equation in $s-$waves to be studied
for binding is, for $\hat \varphi \equiv x \varphi,$
$$
\left( -{d^2 \over dx^2} + { (1+\varepsilon/2) \over 2\varepsilon } w(x) +
{ c \gamma \over 4 (1+c) x} - {\Lambda \over x^2} -
{ (1+\varepsilon/2) E  \over 2\varepsilon } \right)
\hat \varphi=0, \ \ x \ge 0, \ \  \hat \varphi(0)=0, \eqno(3.21)
$$
with
$$
\Lambda=-{ (1+\varepsilon/2) c^2 \over 2\varepsilon } - {1 \over 4} +
{ 12 + 36 c + 45 c^2 + c^3 - 13 c^4 - 2 c^5
\over 24 (1 +c)^3 }. \eqno(3.22)
$$
Potentials of order $1/x^3$ when $x \rightarrow \infty$ do not
modify our expected conclusion of a large number of bound states and can be
simply neglected. We also keep in mind that the $1/x$ and $1/x^2$
forms contributing to Eq.(3.21) are not valid for small values of $x,$
since the potentials they represent are actually finite.

\medskip
If now we assume for the sake of simplicity
that $w$ has a strictly finite range
$x_0,$ the radial equation Eq.(3.21) can be replaced by
$$
\left( -{d^2 \over dx^2} +
{ c \gamma \over 4 (1+c) x} - {\Lambda \over x^2} -
{ (1+\varepsilon/2) E \over 2\varepsilon } \right)
\hat \varphi=0, \ \ x \ge x_0, \ \  \hat \varphi(x_0)=0. \eqno(3.23)
$$
The hard core
boundary condition $\hat \varphi(x_0)=0$
removes all effects of the (finite) potentials at short distances.
This introduction of a hard core
is legitimate, for it can only underestimate the number of bound states.

\medskip
A rescaling $x=4(1+c)z/(c\gamma)$ is then suitable to remove the coefficient
of the $1/x$ term. The ratio between $x$ and $z$
is a fixed number, which depends only upon $c$ and $\gamma,$ and we define a
fixed, rescaled hard core radius, $z_0=c\gamma x_0/[4(1+c)],$ accordingly.
Finally the energy scales according to
$E'=4(1+c)^2(2+\varepsilon)E/(c^2\gamma^2\varepsilon),$ and
we find it strictly equivalent to study the number of bound states for
the following problem,
$$
\left( -{d^2 \over dz^2} + {1 \over z+z_0} - {\Lambda \over (z+z_0)^2}
- E' \right) \hat \varphi=0, \ \ z \ge 0, \ \  \hat \varphi(0)=0. \eqno(3.24)
$$

\medskip
The positive coefficient $\Lambda$ is of order
$1/\varepsilon$ and becomes infinite when $\varepsilon \rightarrow 0.$
It is obvious that, as long as $z+z_0 \le \Lambda/2,$ then
$$
{1 \over z+z_0} - {\Lambda \over (z+z_0)^2} \le - {\Lambda \over 2 (z+z_0)^2}.
\eqno (3.25)
$$
Therefore the problem,
$$
\left( -{d^2 \over dz^2} - {\Lambda \over 2 (z+z_0)^2}
- E' \right) \hat \varphi=0, \ \ 0 \le z \le \Lambda/2, \ \
\hat \varphi(0)=\hat \varphi(\Lambda/2)=0, \eqno(3.26)
$$
generates a further lower bound for the number of bound states.
{}From the potential present in Eq.(3.26), we finally use the
trace formula$^{13)}$ for an estimate of a number of bound states,
$$
N_B=\int^{\Lambda/2} dz \ z \ {\Lambda \over 2  (z+z_0)^2 }
\ \propto \ \Lambda \ {\rm Log}(\Lambda). \eqno(3.27)
$$
This number $N_B$ does diverge when $\varepsilon \rightarrow 0.$
\ {\it Q E D}

\medskip
To summarize this section, the Born-Oppenheimer
method is restated as a special case of the
generator coordinate method. The latter is
a projection of the Schroedinger equation onto a trial subspace. The
proof that there is an arbitrarily large number of bound states in
a subspace is extended to the full space of wave functions, by means of the
Hylleraas-Undheim theorem. In the process of relating BO to GCM, however,
a long range repulsive correction $\varepsilon u$
is found to perturb the static
BO potential $\eta$ which creates the Efimov effect. Small values of the
mass ratio, the adiabaticity parameter $\varepsilon,$
are then necessary to validate
our proof of an arbitrarily large number of bound states
for the three-body system.

\ \

\bigskip
\centerline{4. Introduction of Distortions and Mean Field Approximation}

\bigskip
In order to make less mandatory the restriction of $\varepsilon$ to
small values, we suggest a more flexible formulation of the BO method.
For this, we complete $H,$ Eq.(2.1), by a constraint upon $<\vec x>,$ via
an auxiliary, external harmonic oscillator potential $(\vec x-\vec \xi)^2$ with
a strong spring constant, Lagrange multiplier $L.$ This defines the constrained
Hamiltonian
$$
{\cal H}= P^2/M + L\ (\vec x-\vec \xi)^2 + w + p^2 + \lambda
(TvT^{-1}+T^{-1}vT).
\eqno(4.1)
$$
It is clear that the ground state
$|\Phi_{\vec\xi}>$ of ${\cal H}$ verifies
$<\Phi_{\vec \xi}|\vec x|\Phi_{\vec \xi}> \rightarrow \vec\xi$ when
$L \rightarrow \infty.$ For finite values of $L,$ however, the freezing
of $\vec x$ is implemented by a wave packet, less stringent than
a $\delta-$function $\delta(\vec x-\vec \xi).$

\medskip
For simplicity of notations
in the following, vectors will again be replaced by scalars.
The simplest approximate description of the ground state of ${\cal H}$
is obtained by means of a Hartree ansatz
$\Phi_{\xi}(x,y)=\Gamma_{\xi}(x)\ \Xi_{\xi}(y).$ It is clear that $\Gamma$
is not very different from a sharp Gaussian centered at $\xi,$ and that,
however, it incorporates distortions due to the various potentials present
into the corresponding Hartree equation,
$$
\left[P^2/M+L\ (x-\xi)^2+w+{\cal V}-\eta'\right]\ \Gamma=0, \eqno(4.2)
$$
where ${\cal V}$ is the mean field potential induced
by the convolution of $\lambda(TvT^{-1} + T^{-1}vT)$ with the density
$[\Xi(y)]^2.$
The corresponding Hartree eigenvalue $\eta'$
contains a spurious harmonic oscillator contribution, to be discarded
if physical interpretations are needed.
In turn, $\Xi$ is the bound state generated for the third
particle according to the second Hartree equation,
$$
(p^2 + {\cal U}_{\xi} - \eta'')\ \Xi=0, \eqno(4.3)
$$
where ${\cal U}_{\xi}$ is
the mean field potential arising from the convolution
of $\lambda(TvT^{-1} + T^{-1}vT)$
with the density $[\Gamma(x)]^2.$ In so far as
$\Gamma$ is strongly localized around $\xi,$ there is not much difference
between ${\cal U}_{\xi}$ and $\lambda [v(y-\xi/2)+v(y+\xi/2)].$ Hence there
is a strong similarity between this second Hartree equation,
Eq.(4.3) and the fast BO equation driven by $h_f,$ see Eq.(2.7).
The same similarity holds for $\eta''(\xi)$ and the static BO potential
$\eta(\xi).$ The former, however, contains those corrections arising
from the differences between $\Gamma$ and a $\delta-$function. It is stressed
that all these corrections and distorsions represent self consistency
between $x$ and $y,$ and facilitate binding.

\medskip
The generator coordinate ansatz,
$$
|\Psi>=\int d\xi \ F(\xi) \ |\Phi_{\xi}>, \eqno(4.4)
$$
then leads to the usual GHW equation,
$$
\int d\xi\
\left(<\Phi_{\xi'}|H|\Phi_{\xi}>-E<\Phi_{\xi'}|\Phi_{\xi}> \right)\ F(\xi)=0,
\eqno(4.5)
$$
where now the overlap kernel
${\cal N}_{\xi'\xi} \equiv <\Phi_{\xi'}|\Phi_{\xi}>$ differs from
$\delta(\xi-\xi').$ It is rather similar to a Gaussian, as an overlap of
the quasi-Gaussian wave packets $\Gamma_{\xi}$  and $\Gamma_{\xi'}.$ Its
inverse matrix square root ${\cal N}^{-1/2}$ will be necessary for
the usual GCM deconvolution manipulation. The slow wave function will not be
$F,$ but rather ${\cal N}^{1/2}F.$

\medskip
It may then be
convenient to write the Hamiltonian
kernel $\hat H_{\xi'\xi} \equiv <\Phi_{\xi'}|H|\Phi_{\xi}>$ under the form
$$
\eqalign {
&\hat H_{\xi'\xi}=<\Phi_{\xi'}|\left[P^2/M+w+p^2+{\cal U}_{\xi}
+\lambda(TvT^{-1} + T^{-1}vT)-{\cal U}_{\xi}\right]|\Phi_{\xi}>\cr
&=<\Phi_{\xi'}|(P^2/M+w)|\Phi_{\xi}>+
<\Phi_{\xi'}|[\lambda(TvT^{-1} + T^{-1}vT)-{\cal U}_{\xi}]|\Phi_{\xi}>
+ \eta''(\xi){\cal N}_{\xi'\xi},
}, \eqno(4.6)
$$
for a hint that $\eta''$ will appear as a dynamical effective potential
when the usual deconvolution ${\cal N}^{-1/2} \hat H {\cal N}^{-1/2}$
is performed. It will be noticed that,
according to the Hartree definition of ${\cal U},$ the
diagonal matrix element
$<\Phi_{\xi}|[\lambda(TvT^{-1} + T^{-1}vT)-{\cal U}_{\xi}]|\Phi_{\xi}>$
vanishes identically. Finally it is known$^{12)}$ that the same deconvolution
removes the zero-point kinetic energy which plagues
$<\Phi_{\xi'}|P^2/M|\Phi_{\xi}>.$

\medskip
All told, after deconvolution, the present GHW equation, which includes
dynamical distortions, reads as a generalization of the slow BO equation.
The self consistency inserted by the Hartree method,
via those distortions included in $\Gamma$
and $\Xi,$ is expected to bring more binding. It should thus increase
the number of bound states and allow larger values of $\varepsilon$
to be compatible with a given, large number of bound states.
%
%
%

\ \

\bigskip
\centerline{5. 
Discussion and Conclusion}

\bigskip
In this paper, on one hand, we provide another {
rigorous}
proof of an Efimov effect. But the proven effect is weaker than the full
expected effect. Namely, given the initial hypothesis that a pair
potential is marginally able to bind, an arbitrarily large number of three-body
bound states is obtained only if a suitable mass ratio is small enough. This
may be the result of the BO ansatz, Eq.(3.1), and the particular coordinate
system chosen.

\medskip
On the other hand, several new results were found. For one,
the scheme of our proof is based on arguments of moderate technicality only,
such as the recognition that the BO method is but a special
case of a projection of the three-body dynamics
into a generator coordinate subspace. Hence, we were
able to take advantage of the Hylleraas-Undheim theorem, which states that
there are at least as many bound states in the full space of wave functions as
there are in a subspace.

\medskip
We were also able to relax the BO freezing of the heavy
degree of freedom into more flexible Hartree calculations under
harmonic oscillator constraints. This allows more precise descriptions of the
wave functions of Efimov states, including mean field distortions.

\medskip
Despite the slight technical difficulties we had
to face when the effective potential, of the form $-1/x^2,$ became plagued by
corrections of the form $\varepsilon/x,$ we did not find it necessary here
to introduce the now well known electron translation factor (ETF)
correction of the BO wave functions for the calculation
of the molecular potential$^{14-17)}$. An introduction of this ETF
correction is likely to be in order for future stages of the theory only. We
conjecture that the $\varepsilon/x$ perturbation may be eliminated when
Eq.(3.1) is improved by this ETF asymptotically. The problem of ill-behaved
boundary conditions raised by the conflict between the representation
where the two heavy particles are considered close by and the representation
where these heavy particles are taken far apart is not new and smooth
transitions between such representations have been proposed$^{16)}.$
We can point out, however, that each one of the two competing representations
provides a complete basis of the Hilbert space of bound states.
For our theory of Efimov bound states, the conflict between these
representations is thus tempered.

\medskip
Accordingly, we find it reasonable, while odd at first sight,
that the reduced mass which we use to define the
critical value of the potential strength is a three-body reduced mass
$\mu=2Mm/(2M+m)$ rather than a pair reduced mass $\mu'=Mm/(M+m).$ This choice
is indeed imposed mathematically by the ``fast'' BO
Hamiltonian, see Eq.(2.7). But it may also receive an intuitive, physical
interpretation : that BO bound state $\chi_x(y),$ which reaches zero
binding when $x \rightarrow \infty,$ is even under the exchange of the two
heavy particles and is defined with respect to a fictitious particle,
namely the center of mass of these heavy particles. In that sense,
the critical condition for marginal binding must refer to
the relation between the light particle and that center of mass with mass $2M.$
All told, for Efimov states, where long range effects are at work, a pair
formed by the light particle and one of the heavy ones cannot really be
isolated from the other heavy particle.

\medskip
Finally, our approach can be extended$^{18)}$ by the consideration of coupling
the three possible ``channels'' defined by the three possible pair partitions.
A BO treatment, or the constrained Hartree(-Fock)
generalization advocated in Section 4, can be undertaken for each such channel.
There is no doubt that the resulting, projected Faddeev equations will be
driven by long range potentials similar to those found in the one-channel,
mathematically rigorous argument detailed in Sections 2 and 3.

\medskip 

\noindent
{\it Acknowledgements :}  One author (YH) thanks the
Saclay theory group for their hospitality during three visits
since 1990. The other (BG) is grateful to the University of Connecticut for
a visit, where this work was started, and thanks J.Letourneux
for pointing out an inconsistency in an earlier version of the manuscript.

\medskip 

\noindent \centerline{{\bf References}}

\medskip \noindent
1)V.Efimov, Yad.Fiz.{\bf 12},1080(1970)[Sov.J.Nucl.Phys.{\bf 12},589(1971)];
Phys.Lett.{\bf 33B},563 \par
\noindent
(1970); Nucl.Phys.{\bf A210},579(1973)

\medskip \noindent
2)R.D.Amado and J.V.Noble, Phys.Lett.{\bf 35B},25(1971);
Phys.Rev.{\bf D5},1992(1972)

\medskip \noindent
3)A.C.Fonseca and P.E.Shanley, Ann.Phys.(NY){\bf 117},268(1979)

\medskip \noindent
4)A.C.Fonseca, E.F.Redish and P.E.Shanley, Nucl.Phys.{\bf A320},273(1979)

\medskip \noindent
5)Yu.N.Ouchinnikov and I.M.Segal, Ann.Phys.(NY){\bf 123},274(179)

\medskip \noindent
6)R.D.Amado and F.C.Greenwood, Phys.Rev.{\bf D7},2517(1973)

\medskip \noindent
7)H.Kruger and R.Perne, Phys.Rev.{\bf C22},21(1980)

\medskip \noindent
8)S.K.Adhikari and A.C.Fonseca, Phys.Rev.{\bf D24},416(1981)

\medskip \noindent
9)R.G.Newton, {\it Scattering Theory of Waves and Particles,} 2nd ed.,
Springer-Verlag(1982), pp.324-327

\medskip \noindent
10)D.L.Hill and J.A.Wheeler, Phys.Rev.{\bf 89},1102(1953)

\medskip \noindent
11)J.J.Griffin and J.A.Wheeler, Phys.Rev.{\bf 108},311(1957)

\medskip \noindent
12)B.Giraud, J.Letourneux and E.Osnes, Ann.Phys.(NY){\bf 89},359(1975)

\medskip \noindent
13)R.G.Newton, {\it Scattering Theory of Waves and Particles,} 2nd ed.,
Springer-Verlag(1982), p.358

\medskip \noindent
14)D.R.Bates and R.MacCarroll, Proc.Roy.Soc.{\bf A245},175(1958)

\medskip \noindent
15)Y.Hahn, Phys.Rev.{\bf 154},981(1961)

\medskip \noindent
16)A.V. Matveenko and Y.Abe, Few-Body Systems,{\bf 2},127(1987)

\medskip \noindent
17)C.J. Kleinman, Y. Hahn and L. Spruch, Phys.Rev.{\bf 165},53(1968)

\medskip \noindent
18)A.C. Fonseca and M.T. Pe\~na, Phys.Rev.{\bf A36},4585(1987);
Phys.Rev.{\bf A38},4967(1988)

\bye